\documentclass[intlimits,twoside,a4paper]{article}

\usepackage{amsmath,amssymb}

\usepackage[T2A]{fontenc}
\usepackage[cp1251]{inputenc}
\usepackage{cmpj2}

\articletype{Rapid Communication}

\issue{2012}{15}{1}{14001}

\doinumber{10.5488/CMP.15.14001}

\title[On the problem of a consistent description...]%
{ On the problem of a consistent description of kinetic and
hydrodynamic processes in dense gases and liquids: Collective
excitations spectrum }
\author[B.B. Markiv, I.P. Omelyan, M.V. Tokarchuk]
{B.B. Markiv, I.P. Omelyan, M.V. Tokarchuk}
\address{Institute for Condensed Matter Physics of the
National Academy of Sciences of Ukraine, \\1 Svientsitskii Str.,
79011 Lviv, Ukraine}
\date{Received December 21, 2011, in final form March 5, 2012}

\authorcopyright{B.B. Markiv, I.P. Omelyan, M.V. Tokarchuk, 2012}

\begin{document}

\maketitle

\begin{abstract}
Based on the generalized non-Markovian
equations obtained earlier for a nonequilibrium one-particle distribution
function and potential part of the averaged enthalpy density
[Markiv~B.B., Omelyan~I.P, Tokar\-chuk~M.V., Condens. Matter
Phys., 2010, \textbf{13}, 23005] a spectrum of collective excitations
is investigated, where the potential of interaction between particles is
presented as a sum of the potential of hard spheres and a certain
long-range potential.

\pacs 05.20.Dd, 05.60.+w, 52.25.Fi, 82.20.M
\end{abstract}

\section{Introduction}

A number of investigations~\cite{118,77,zub5,zub6,klim2,tok2,zub4,tok,zub2,mar}
were devoted to the problem of constructing a consistent
description of kinetic and hydrodynamic processes in dense gases,
liquids, and plasma. For instance, the
importance of taking into account the kinetic processes connected
with irreversible collision processes at the scale of short-ranged
interparticle interactions was pointed out in~\cite{deSchep}. Short-wavelength collective modes in liquids were
investi\-gated therein based on the linearized kinetic
equation of the revised Enskog theory for the hard
spheres model.

In this paper we investigate a spectrum of collective excitations
within a consistent description of kinetic and hydrodynamic
processes in a system in which the potential of interaction between
particles consists of two parts: the hard spheres potential and a
long-range part.

\section{Transport equations}

Using the ideas presented in papers~\cite{zub5,zub6} the nonequilibrium
statistical operator consistently describing the kinetic and
hydrodynamic processes for a system of classical interacting
particles was obtained in~\cite{zub4,tok} by means of
nonequilibrium statistical operator method. Using this operator,
a set of kinetic equations for the nonequilibrium one-particle
distribution function
$f_{\vec{k}}(\vec{p};t)=\langle\hat{n}_{\vec{k}}(\vec{p})\rangle^t$
and the potential part of the averaged enthalpy density
$h^\mathrm{int}_{\vec{k}}(t)=\langle\hat{h}^\mathrm{int}_{\vec{k}}\rangle^t$ was
obtained in the case of weakly nonequilibrium processes:
\begin{eqnarray}
\label{math/2.11} \lefteqn{\frac{\partial}{\partial
t}f_{\vec{k}}(\vec{p};t)+\frac{\ri\vec{k}\cdot\vec{p}}{m}f_{\vec{k}}(\vec{p};t)
=-\frac{\ri\vec{k}\cdot\vec{p}}{m}nf_0(p)c_2(k)\int
\rd\vec{p}'f_{\vec{k}}(\vec{p}';t)+\ri\Omega_{nh}(\vec{k};\vec{p})h^\mathrm{int}_{\vec{k}}(t)}\nonumber
\\&&\mbox{}-\int
\rd\vec{p}'\int_{-\infty}^{t}\re^{\varepsilon(t'-t)}
\varphi_{nn}(\vec{k};\vec{p},\vec{p}';t,t')f_{\vec{k}}(\vec{p}';t')\rd t'
-\int_{-\infty}^{t}\re^{\varepsilon(t'-t)}
\varphi_{nh}(\vec{k};\vec{p};t,t')h^\mathrm{int}_{\vec{k}}(t')\rd t',
\end{eqnarray}
\begin{eqnarray}
\label{math/2.12} \lefteqn{\frac{\partial}{\partial
t}h_{\vec{k}}^\mathrm{int}(t)=\int
\rd\vec{p}'\ri\Omega_{hn}(\vec{k};\vec{p}')f_{\vec{k}}(\vec{p}';t)}\nonumber
\\&&\mbox{}-\int
\rd\vec{p}'\int_{-\infty}^{t}\re^{\varepsilon(t'-t)}
\varphi_{hn}(\vec{k};\vec{p}';t,t')f_{\vec{k}}(\vec{p}';t')\rd t'
-\int_{-\infty}^{t}\re^{\varepsilon(t'-t)}
\varphi_{hh}(\vec{k};t,t')h^\mathrm{int}_{\vec{k}}(t')\rd t',
\end{eqnarray}
where $\ri\Omega_{nh}(\vec{k};\vec{p})=
\langle\dot{\hat{n}}_{\vec{k}}(\vec{p})\hat{h}^\mathrm{int}_{-\vec{k}}\rangle_0
\Phi_{hh}^{-1}(\vec{k})$ and
$\ri\Omega_{hn}(\vec{k};\vec{p})= \int
\rd\vec{p}'\langle\dot{\hat{h}}_{\vec{k}}^\mathrm{int}\hat{n}_{-\vec{k}}(\vec{p}')\rangle_0
\Phi_{\vec{k}}^{-1}(\vec{p}',\vec{p})$ are the normalized static
correlation functions.
\begin{eqnarray}\label{math/2.14}
\varphi_{nn}(\vec{k};\vec{p},\vec{p}';t,t')&=& \int
\rd\vec{p}''\langle
I_{n}(\vec{k};\vec{p})T_0(t,t')I_{n}(-\vec{k};\vec{p}'')\rangle_0
\Phi_{\vec{k}}^{-1}(\vec{p}'',\vec{p}'),\nonumber\\
\varphi_{hn}(\vec{k};\vec{p};t,t')&=& \int \rd\vec{p}'\langle
I_{h}^\mathrm{int}(\vec{k})T_0(t,t')I_{n}(-\vec{k};\vec{p}')\rangle_0
\Phi_{\vec{k}}^{-1}(\vec{p}',\vec{p}),\nonumber\\
\varphi_{nh}(\vec{k};\vec{p};t,t')&=&\langle
I_{n}(\vec{k};\vec{p})T_0(t,t')I_{h}^\mathrm{int}(-\vec{k})\rangle_0
\Phi_{hh}^{-1}(\vec{k})\nonumber,\\
\varphi_{hh}(\vec{k};t,t')&=&
\langle
I_{h}^\mathrm{int}(\vec{k})T_0(t,t')I_{h}^\mathrm{int}(-\vec{k})\rangle_0
\Phi_{hh}^{-1}(\vec{k})
\end{eqnarray}
are the generalized transport kernels
(memory functions) describing kinetic and hydrodynamic processes.
In~\cite{marometok}, the inner structure of generalized transport
kernels for a consistent description of kinetic and hydrodynamic
processes was analyzed in detail. It was shown that they are
expressed in terms of time correlation functions related to the
basic set of dynamical variables, phase density
$\hat{n}_{\vec{k}}(\vec{p})$ and potential part of the enthalpy
density $\hat{h}_{\vec{k}}^\mathrm{int}$ along with the transport
kernels describing  diffusive and visco-thermal processes. Here,
$\hat{n}_{\vec{k}}(\vec{p})=\int
\rd\vec{r}\re^{-\ri\vec{k}\vec{r}}\hat{n}_1(\vec{r},\vec{p})$ are the
Fourier-components of microscopic phase density of particles
number,
$\hat{n}_1(\vec{r},\vec{p})=\sum_{l=1}^N\delta(\vec{p}-\vec{p}_l)
\delta(\vec{r}-\vec{r}_l)$,
$\hat{h}_{\vec{k}}^\mathrm{int}=\hat{\varepsilon}_{\vec{k}}^\mathrm{int}
-\langle\hat{\varepsilon}_{\vec{k}}^\mathrm{int}\hat{n}_{-\vec{k}}\rangle_0S^{-1}(k)
\hat{n}_{\vec{k}}$ are the Fourier-components of the potential
part of the enthalpy density,
$\hat{\varepsilon}_{\vec{k}}^\mathrm{int}=\frac{1}{2}\sum_{l\neq
j=1}^N\Phi(|\vec{r}_{lj}|)\re^{-\ri\vec{k}\vec{r}_j}$ and
$\hat{n}_{\vec{k}}=\sum_{l=1}^N\re^{-\ri\vec{k}\vec{r}_l}$ are the
Fourier-components of the potential energy and particle number
densities, respectively, $\vec{k}$ is the wave-vector.
$\Phi_{hh}^{-1}(\vec{k})$ is the function inverse to the
equilibrium correlation function
$\Phi_{hh}(\vec{k})=\langle\hat{h}_{\vec{k}}^\mathrm{int}
\hat{h}_{-\vec{k}}^\mathrm{int}\rangle_0,$
$\langle\ldots\rangle_0=\int \rd\Gamma_N\ldots\varrho_0(x^N)$, where
$\varrho_0$ is an equilibrium statistical operator.
$I_n(\vec{k};\vec{p})=(1-P_0)\ri L_N\hat{n}_{\vec{k}}(\vec{p})
=(1-P_0)\dot{\hat{n}}_{\vec{k}}(\vec{p})$ and
$I_h^\mathrm{int}(\vec{k})=(1-P_0)\ri L_N\hat{h}^\mathrm{int}_{\vec{k}}
=(1-P_0)\dot{\hat{h}}^\mathrm{int}_{\vec{k}}$ are the generalized flows
in linear approximation, $T_0(t,t')=\re^{(t-t')(1-P_0)\ri L_N}$ is the
evolution operator with regard to projection. $P_0$
is the linear approximation of the Mori projection operator
constructed on the orthogonal dynamic variables
$\hat{n}_{\vec{k}}(\vec{p})$,
$\hat{h}_{\vec{k}}^\mathrm{int}$~\cite{tok}:
$P_0\hat{A}_{\vec{k}}={\sum_{\vec{k}}}
\langle\hat{A}_{\vec{k}}\hat{h}^\mathrm{int}_{-\vec{k}}\rangle_0\Phi_{hh}^{-1}(\vec{k})
\hat{h}^\mathrm{int}_{\vec{k}}+{\sum_{\vec{k}}}\int \rd\vec{p}\int
\rd\vec{p}'
\langle\hat{A}_{\vec{k}}\hat{n}_{-\vec{k}}(\vec{p})\rangle_0
\Phi_{\vec{k}}^{-1}(\vec{p},\vec{p}')
\hat{n}_{\vec{k}}(\vec{p}')$.
It possesses the following properties: $P_0P_0=P_0$, $P_0(1-P_0)=0$,
$P_0\hat{n}_{\vec{k}}(\vec{p})=\hat{n}_{\vec{k}}(\vec{p})$,
$P_0\hat{h}^\mathrm{int}_{\vec{k}}=\hat{h}^\mathrm{int}_{\vec{k}}$.
$\Phi_{\vec{k}}^{-1}(\vec{p},\vec{p}')$ is the function inverse to
$\Phi_{\vec{k}}(\vec{p},\vec{p}')=\langle\hat{n}_{\vec{k}}(\vec{p})
\hat{n}_{-\vec{k}}(\vec{p}')\rangle_0=n\delta(\vec{p}-\vec{p}')f_0(p')
+n^2f_0(p)f_0(p')h_2(\vec{k})$. It is equal to
$\Phi_{\vec{k}}^{-1}(\vec{p},\vec{p}')=\frac{\delta(\vec{p}-\vec{p}')}{nf_0(p')}-c_2(k)$,
where $n=N/V$, $f_0(p)=\left({\beta}/{2\pi
m}\right)^{3/2}\re^{-\beta\frac{p^2}{2m}}$ is the Maxwellian
distribution, $\beta=1/k_\mathrm{B}T$ is an inverse temperature and $k_\mathrm{B}$
is Boltzmann constant. $c_2(k)$ is the direct correlation function
related to the correlation function $h_2(k)$:
$h_2(k)=c_2(k)[1-nc_2(k)]^{-1}$.
$S(k)=\langle\hat{n}_{\vec{k}}\hat{n}_{-\vec{k}}\rangle_0$ denotes the
static structure factor. It is important to note that dynamical
variables $\hat{h}_{\vec{k}}^\mathrm{int}$ and
$\hat{n}_{\vec{k}}(\vec{p})$ are orthogonal in the sense that
$\langle\hat{h}_{\vec{k}}^\mathrm{int}\hat{n}_{\vec{k}}(\vec{p})\rangle_0=0$.

Projecting the set of equations (\ref{math/2.11}),
(\ref{math/2.12}) onto the first moments of the nonequilibrium
one-particle distribution function $\Psi_1(\vec{p})=1$,
$\Psi_{\alpha}(\vec{p})=\sqrt{2}p_{\alpha}/2k_\mathrm{B}T$ (where
$\alpha=x,y,z$),
$\Psi_{\varepsilon}(\vec{p})=\sqrt{2/3}(p^2/2mk_\mathrm{B}T-3/2)$, one can
obtain a set of equations for the averaged values of densities
of particles number $n_{\vec{k}}(t)$, momentum
$\vec{\jmath}_{\vec{k}}(t)$, kinetic $h_{\vec{k}}^\mathrm{kin}(t)$ and
potential $h_{\vec{k}}^\mathrm{int}(t)$ parts of enthalpy~\cite{tok}, where the Fourier-components of the kinetic part of enthalpy density defined as $\hat{h}_{\vec{k}}^\mathrm{kin}=\hat{\varepsilon}_{\vec{k}}^\mathrm{kin}
-\langle\hat{\varepsilon}_{\vec{k}}^\mathrm{kin}\hat{n}_{-\vec{k}}\rangle_0
\langle\hat{n}_{\vec{k}}\hat{n}_{-\vec{k}}\rangle_0^{-1}\hat{n}_{\vec{k}}$. For
this purpose, we introduce the projection operator constructed on
the eigenfunctions $|\Psi_{\nu}(\vec{p})\rangle$ of the
nonequilibrium one-particle function such that ${\cal
P}|\Psi\rangle=\sum_{\nu=1}^n|\Psi_{\nu}\rangle\langle\Psi_{\nu}|\Psi\rangle$.
Here, $\langle\Psi|\Psi_{\nu}\rangle=\int
d\vec{p}\Psi(\vec{p})f_0(p)\Psi_{\nu}(\vec{p})$, while
$\Psi_{\nu}(\vec{p})$ satisfies the conditions
$\langle\Psi_{\mu}|\Psi_{\nu}\rangle=\delta_{\mu\nu}$ and
$\sum_{\nu}|\Psi_{\nu}\rangle\langle\Psi_{\nu}|=1$. Then, let us
act by the projection operator $\cal P$ onto the set of equation
(\ref{math/2.11}), (\ref{math/2.12}). Repeat this operation acting
by the operator ${\cal Q}=1-{\cal P}$ complementary to ${\cal P}$.
Then, substituting the unknown quantity from the second equation
into the first one we obtain the necessary set of equations with separated contributions of kinetic and potential energies. Using the
Laplace transform, let us represent it in a matrix form:
\begin{eqnarray}
\label{math/3.432}
z\tilde{a}_{\vec{k}}(z)-\tilde{\Sigma}_\mathrm{G}(\vec{k};z)\tilde{a}_{\vec{k}}(z)=-\langle
\tilde{a}_{\vec{k}}(t=0)\rangle^{t}.
\end{eqnarray}
$\tilde{\Sigma}_\mathrm{G}(\vec{k};z)$ is the matrix of memory kernels
\begin{eqnarray}
\label{math/3.44} \tilde{\Sigma}_\mathrm{G}(\vec{k};z)
=\ri\tilde\Omega_\mathrm{G}(\vec{k})-\tilde\Pi(\vec{k};z),
\end{eqnarray}
where $\tilde{a}_{\vec{k}}(z)=[n_{\vec{k}}(z),
\vec{\jmath}_{\vec{k}}(z), h_{\vec{k}}^\mathrm{kin}(z),
h_{\vec{k}}^\mathrm{int}(z)]$ is the column-vector.
\begin{eqnarray}
\label{math/3.46} \ri\tilde{\Omega}_\mathrm{G}(\vec{k})=\left(
\begin{array}{llll}
  0       & \ri\Omega_{n\jmath} & 0 & 0 \\
  \ri\Omega_{\jmath n} & 0 & \ri\Omega_{\jmath h}^\mathrm{kin} & \ri\Omega_{\jmath h}^\mathrm{int} \\
  0 & \ri\Omega_{h\jmath}^\mathrm{kin} & 0 & 0 \\
  0 & \ri\Omega_{h\jmath}^\mathrm{int} & 0 & 0 \\
\end{array}\right),
\quad
\tilde{\Pi}(\vec{k};z)=\left(
\begin{array}{llll}
  0 & 0 & 0 & 0 \\
  0 & \Pi_{\jmath\jmath} & \Pi_{\jmath h}^\mathrm{kin} & \Pi_{\jmath h}^\mathrm{int} \\
  0 & \Pi_{h\jmath}^\mathrm{kin} & \Pi_{hh}^\mathrm{kin,kin} & \Pi_{hh}^\mathrm{kin,int} \\
  0 & \Pi_{h\jmath}^\mathrm{int} & \Pi_{hh}^\mathrm{int,kin} & \Pi_{hh}^\mathrm{int,int} \\
\end{array}\right)
\end{eqnarray}
are the frequency matrix and
the matrix of transport kernels. The elements of the latter have
the following structure:
\begin{eqnarray}
\label{math/3.47} \Pi_{\mu\nu}(\vec{k};z)
=\langle\Psi_{\mu}|\tilde{\varphi}(\vec{k};z)+\tilde{\Sigma}(\vec{k};z)
{\cal Q}[z\tilde{I}-{\cal Q}\tilde{\Sigma}(\vec{k};z){\cal
Q}]^{-1}{\cal Q}\tilde{\Sigma}(\vec{k};z)|\Psi_{\nu}\rangle.
\end{eqnarray}
$\tilde{I}$ denotes a unit matrix, $\tilde{\varphi}(\vec{k};z)$ is the matrix whose elements
are the generalized transport kernels \linebreak
$\varphi_{nn}(\vec{k};\vec{p},\vec{p}';z)$,
$\varphi_{hn}(\vec{k};\vec{p}';z)$,
$\varphi_{nh}(\vec{k};\vec{p};z)$, $\varphi_{hh}(\vec{k};z)$ in the set of equations (\ref{math/2.11}), (\ref{math/2.12}), and
$\tilde{\Sigma}(\vec{k};z)=\ri\tilde{\Omega}(\vec{k})-
\tilde{\varphi}(\vec{k};z)$. Here, $\ri\tilde{\Omega}(\vec{k})$ is
the matrix of static correlation functions
$\ri\Omega_{nh}(\vec{k};\vec{p})$, $\ri\Omega_{hn}(\vec{k};\vec{p})$.
For the sake of simplicity the dependence of $\tilde{\varphi}(\vec{k};z)$, $\tilde{\Sigma}(\vec{k};z)$ on $\vec{p},\vec{p}'$ was omitted.
As we can see from the structure of elements of the matrices
$\ri\tilde\Omega_\mathrm{G}(\vec{k})$ and $\tilde\Pi(\vec{k;z})$, the
contributions of kinetic and potential parts of enthalpy are
separated. Herewith, a question arises regarding the study of time
correlation functions and collective modes for liquids based on the set of transport
equations~(\ref{math/3.432}).

\section{Spectrum of collective excitations}

Let us consider the system of kinetic equations~(\ref{math/2.11}),
(\ref{math/2.12}) in the case where the potential of interaction is
presented as follows:
\begin{eqnarray}
\label{math/3.48} \Phi(|\vec{r}_{ij}|)=\Phi^\mathrm{hs}(|\vec{r}_{ij}|)
+\Phi^\mathrm{l}(|\vec{r}_{ij}|),
\end{eqnarray}
where $\Phi^\mathrm{hs}(|\vec{r}_{ij}|)$ is the hard sphere interaction
potential, and $\Phi^\mathrm{l}(|\vec{r}_{ij}|)$ is the long-range
potential. Taking into account the features of the hard sphere
model dynamics~\cite{zub6} and the results of investigations~\cite{117,114,deSchep}, one can separate Enskog-Boltzmann
collision integral from the function
$\varphi_{nn}(\vec{k};\vec{p},\vec{p}';t,t')$. Indeed, an
infinitesimal time of a collision $\tau_0\rightarrow+0$ within an
infinitesimal region $\sigma\pm\Delta r_0$, $\Delta
r_0\sim|\tau_0||\vec{p}_2-\vec{p}_1|/m\rightarrow+0$ being a
feature of the hard sphere model dynamics ($\sigma$ is the
hard sphere diameter). Taking this into account in the kinetic
equation~(\ref{math/2.11}) we can obtain the kinetic equation of the revised Enskog
theory for the hard sphere model and the kinetic Enskog-Landau
equation for the charged hard sphere model in a pair collision
approximation, respectively~\cite{zub6}. In the latter case, when
$\Phi^\mathrm{l}(|\vec{r}_{ij}|)$ is the Coulomb potential of interaction,
taking into account the features $\tau_0\rightarrow+0$, $\Delta
r_0\rightarrow+0$ makes it possible to separate a collision
integral of the revised Enskog theory and a Landau-like collision
integral in the limits $\tau\rightarrow-0$ and
$\tau\rightarrow-\infty$, respectively. In the case of potential
(\ref{math/3.48}), in the region of $\tau_0\rightarrow+0$, $\Delta
r_0\rightarrow+0$, $\sigma\pm\Delta r_0$ where the main
contribution to a dynamics is defined by pair collisions of hard
spheres, the memory function
$\varphi_{nn}(\vec{k};\vec{p},\vec{p}';t,t')$ can be calculated by
expanding it over the density (a pair collision approximation),
which was scrupulously done in papers by Mazenko~\cite{109,110,111,114,117}.

Then, the kinetic equation~(\ref{math/2.11}) can be represented in
the following form:
\begin{eqnarray}
\label{math/3.49} \lefteqn{\frac{\partial}{\partial
t}f_{\vec{k}}(\vec{p};t)+\frac{{\rm
i}\vec{k}\cdot\vec{p}}{m}f_{\vec{k}}(\vec{p};t)=-\frac{{\rm
i}\vec{k}\cdot\vec{p}}{m}nf_{0}(\vec{p})\left[c_2(k)-g_2(\sigma)c_2^0(k)\right]\int
\rd\vec{p}'f_{\vec{k}}(\vec{p}';t)}\nonumber
\\&&\mbox{}-\int
\rd\vec{p}'
\varphi_{nn}^\mathrm{hs}(\vec{k},\vec{p},\vec{p}')f_{\vec{k}}(\vec{p}';t)
+{\rm i}\Omega_{nh}(\vec{k};\vec{p})h_{\vec{k}}^\mathrm{int}(t) \nonumber
\\&&\mbox{}-\int
\rd\vec{p}'\int_{-\infty}^t \rd t'{\re}^{\varepsilon(t-t')}
\varphi_{nn}^\mathrm{l}(\vec{k};\vec{p},\vec{p}';t,t')f_{\vec{k}}(\vec{p}';t')
-\int_{-\infty}^t \rd t'{\re}^{\varepsilon(t-t')}\varphi_{nh}(\vec{k};\vec{p};t,t')h_{\vec{k}}^\mathrm{int}(t').
\end{eqnarray}
Here,
\begin{eqnarray}
\label{math/3.49a} \lefteqn{\int \rd\vec{p}'
\varphi_{nn}^\mathrm{hs}(\vec{k},\vec{p},\vec{p}')f_{\vec{k}}(\vec{p}';t)=
 ng_2(\sigma)\sigma^2\int \rd\Omega_{\sigma}\int
\rd\vec{p}'\frac{(\vec{p}-\vec{p}')\cdot\hat{\vec{\sigma}}}{m}
\Theta_{-}\left(\hat{\vec{\sigma}}\cdot[\vec{p}-\vec{p}']\right)}\nonumber
\\&&\mbox{}
\times\left[f_0(p'^*)f_{\vec{k}}(\vec{p};t)-f_0(p')f_{\vec{k}}(\vec{p}^*;t) +
{\re}^{{\ri}\vec{k}\cdot\hat{\vec{\sigma}}\sigma}f_0(p'^*)f_{\vec{k}}(\vec{p}'^*;t)-
{\re}^{{\ri}\vec{k}\cdot\hat{\vec{\sigma}}\sigma}f_0(p)f_{\vec{k}}(\vec{p}';t)\right]
\end{eqnarray}
is the Enskog-Boltzmann collision integral, where
$c_2^0(\vec{k})$ is the low-density limit of the direct
correlation function and $g_2(\sigma)$ is the pair distribution
function. The step function $\Theta_-(x)$ is unity for $x<0$ and
vanishes otherwise. $\rd\Omega_\sigma$ is the differential solid
angle, $\hat{\vec{\sigma}}$ is unity vector. The precollision and
postcollision momenta of the colliding hard spheres are denoted as
$(\vec{p},\vec{p}')$ and $(\vec{p}^*,\vec{p}'^*)$, respectively.
$\varphi_{nn}^\mathrm{l}(\vec{k};\vec{p},\vec{p}';t,t')$ is the part of the
transport kernel related to the long-range interaction potential
$\Phi^\mathrm{l}(|\vec{r}_{ij}|)$. Notably, the presented equation contains
the Enskog-Boltzmann collision integral describing short-time
dynamics of the hard sphere model. The collective effects related
to the long-range interactions between particles are described by
the functions ${\rm i}\Omega_{nh}(\vec{k};\vec{p})$,
$\varphi_{nn}^\mathrm{l}(\vec{k};\vec{p},\vec{p}';t,t')$,
$\varphi_{nh}(\vec{k};t,t')$ and by the equation for
$h_{\vec{k}}^\mathrm{int}(t)$. Since the collective modes for the
Enskog-Boltzmann model are well studied~\cite{deSchep}, the
investigation of time correlation functions and collective modes
for the system of particles interacting through the
potential~(\ref{math/3.48}) turns out to be of great interest. In
the case of the hard spheres system, the set of kinetic
equations~(\ref{math/2.12}), (\ref{math/3.49}) reduces to the
Enskog-Boltzmann kinetic equation~\cite{deSchep}.
\begin{eqnarray}
\label{math/3.50} \lefteqn{\frac{\partial}{\partial
t}f_{\vec{k}}(\vec{p};t)+\frac{{\rm
i}\vec{k}\cdot\vec{p}}{m}f_{\vec{k}}(\vec{p};t)=-\frac{{\rm
i}\vec{k}\cdot\vec{p}}{m}nf_{0}(\vec{p})\left[c_2(k)-g_2(\sigma)c_2^0(k)\right]\int
\rd\vec{p}'f_{\vec{k}}(\vec{p}';t)}\nonumber
\\&&\mbox{}-ng_2(\sigma)\sigma^2\int
\rd\Omega_{\sigma}\int
\rd\vec{p}'\frac{(\vec{p}-\vec{p}')\cdot\hat{\vec{\sigma}}}{m}
\Theta_{-}\left(\hat{\vec{\sigma}}\cdot[\vec{p}-\vec{p}']\right)\nonumber
\\&&\mbox{}\times\left[
f_0(p'^*)f_{\vec{k}}(\vec{p};t)-f_0(p')f_{\vec{k}}(\vec{p}^*;t)+
{\rm e}^{{\rm
i}\vec{k}\cdot\hat{\vec{\sigma}}\sigma}f_0(p'^*)f_{\vec{k}}(\vec{p}'^*;t)-
{\rm e}^{{\rm
i}\vec{k}\cdot\hat{\vec{\sigma}}\sigma}f_0(p)f_{\vec{k}}(\vec{p}';t)\right].
\end{eqnarray}
Projecting the
Enskog-Boltzmann equation~(\ref{math/3.50}) onto the first moments of the
nonequilibrium one-particle distribution function a spectrum of
collective excitations for the hard sphere model was obtained
in~\cite{deSchep,deSchep2}.  Herewith, it is important to note
that for the kinetic Enskog-Boltzmann equation we can consider two
typical limits: $k\sigma\ll1$ and $k\sigma\gg1$. In the hydrodynamic limit
($k\sigma\ll1$) the spectrum includes:
\textbf{ heat mode} $z_\mathrm{H}(k)=-D_\mathrm{TE}k^2$, where $D_\mathrm{TE}$ is
    the thermal diffusivity coefficient in the
    Enskog transport theory~\cite{Chap};
\textbf{ two sound modes} with eigenvalues given by
    $z_{\pm}(k)=\pm \ri ck-\Gamma_\mathrm{E}k^2$, where
    $\Gamma_\mathrm{E}$ is the sound damping coefficient and $c$ is the sound velocity
    in the Enskog theory;
\textbf{ two shear modes} with eigenvalues
    given by $z_{\nu_1}(k)=z_{\nu_2}(k)=z_{\nu}(k)=-\nu_\mathrm{E}k^2$,
    $\nu_\mathrm{E}$ is the kinematic viscosity in the Enskog dense gas
    theory.
In the limit $k\sigma\gg1$ the Enskog-Boltzmann collision
integral (\ref{math/3.49a}) is transformed~\cite{deSchep} into the
Lorentz-Boltzmann collision integral which has only one
eigenfunction $\Psi_1(\vec{p})=1$. Consequently, we obtain the
\textbf{ diffusion mode} only with the eigenvalue
    $z_\mathrm{D}(k)=-D_\mathrm{E}k^2$, where $D_\mathrm{E}$ is the self-diffusion coefficient as given by the Enskog dense gas theory.

Let us now project the system of equations~(\ref{math/2.12}),
(\ref{math/3.49}) onto the first moments of the nonequilibrium
one-particle distribution function. Thereafter, we perform simple
transformations consisting in the transition from the set of
equations (\ref{math/3.432}) for averages $n_{\vec{k}}(z),
\vec{\jmath}_{\vec{k}}(z), h_{\vec{k}}^\mathrm{kin}(z),
h_{\vec{k}}^\mathrm{int}(z)$ to the equations of generalized
hydrodynamics for averages
$\tilde{b}_{\vec{k}}(z)=[n_{\vec{k}}(z),
\vec{\jmath}_{\vec{k}}(z), h_{\vec{k}}(z)=h_{\vec{k}}^\mathrm{kin}(z)+
h_{\vec{k}}^\mathrm{int}(z)]$. This permits to correctly define (see
below) the generalized viscosity coefficient via the transport
kernel~(\ref{math/3.60}) and the heat conductivity coefficient via
the transport kernel $\Pi_{hh}(k,z)$. The averages
$\tilde{b}_{\vec{k}}(z)$ satisfy the set of equations
$z\tilde{b}_{\vec{k}}(z)-\tilde{\Sigma}_\mathrm{G}(\vec{k};z)\tilde{b}_{\vec{k}}(z)=-\langle
\tilde{b}_{\vec{k}}(t=0)\rangle^{t}$. In the limit $k\sigma\gg1$, the latter reduces to a single equation of diffusion for $n_{\vec{k}}(z)$ in which the transport kernel
$\Sigma_\mathrm{G}(\vec{k};z)=\langle\Psi_{1}|\varphi_{nn}^\mathrm{L-B}(\vec{k})|\Psi_{1}\rangle$ corresponds to the Lorentz-Boltzmann collision integral~(\ref{math/3.49a}). In the opposite case, when $k\sigma\ll1$, the matrix $\tilde{\Sigma}_\mathrm{G}(\vec{k};z)$ is defined as follows: $\tilde{\Sigma}_\mathrm{G}(\vec{k};z)=
\tilde{\Sigma}_\mathrm{H}(\vec{k};z)=\ri\tilde\Omega_\mathrm{H}(\vec{k})
-\tilde\Pi_\mathrm{H}(\vec{k};z)$,
\begin{eqnarray}
\label{math/3.466}
\tilde{\Sigma}_\mathrm{H}(\vec{k};z)=\left(
\begin{array}{lll}
   0 & \ri\Omega_{n\jmath} & 0 \\
   \ri\Omega_{\jmath n} & -\langle\Psi_{2}|\varphi_{nn}^\mathrm{hs}|\Psi_{2}\rangle-\Sigma^\mathrm{l}_{jj} &\ri\Omega_{\jmath h}-\Pi_{\jmath h} \\
  0 &\ri\Omega_{h\jmath}- \Pi_{h\jmath} & -\langle\Psi_{3}|\varphi_{nn}^\mathrm{hs}|\Psi_{3}\rangle-\Pi^\mathrm{l}_{hh} \\
  \end{array}\right)_{(k,z)}.
\end{eqnarray}
Here we use the notations
\begin{eqnarray}
\label{math/3.60}
\Sigma_{jj}(k,z)&=&\Pi_{jj}(k,z)-\Sigma_{jh}^\mathrm{int}(k,z) \left[\Sigma_{hh}^\mathrm{kin,kin}(k,z)\right]^{-1}
\Big\{\ri\Omega_{hj}^\mathrm{kin}(k)+\Pi_{hh}^{kin,int}(k,z)\\
&\times& \left[z-\Sigma_{hh}^\mathrm{int,int}(k,z)\right]^{-1} \Sigma_{hj}^\mathrm{int}(k,z)\Big\}-\Sigma_{jh}^\mathrm{kin}(k,z) \left[z-\Sigma_{hh}^\mathrm{int,int}(k,z)\right]^{-1}\Sigma_{hj}^\mathrm{int}(k,z)\nonumber,\\
\label{math/3.61}
\Pi_{hh}(k,z)&=&\Pi_{hh}^\mathrm{kin,kin}(k,z)+\Pi_{hh}^\mathrm{kin,int}(k,z)
        +\Pi_{hh}^\mathrm{int,kin}(k,z)+\Pi_{hh}^\mathrm{int,int}(k,z),
\end{eqnarray}
where  $\Sigma_{hh}^\mathrm{kin,kin}(k,z)=z-\Pi_{hh}^\mathrm{kin,kin}(k,z)$,
$\Sigma_{hh}^\mathrm{int,int}(k,z)=\Pi_{hh}^\mathrm{int,int}(k,z)
+\Pi_{hh}^\mathrm{int,kin}(k,z)\left[\Sigma_{hh}^\mathrm{kin,kin}(k,z)\right]^{-1}\linebreak \times
\Pi_{hh}^\mathrm{kin,int}(k,z)$, \
$\Sigma_{hj}^\mathrm{int}(k,z)=\ri\Omega_{hj}^\mathrm{int}(k)+\Pi_{hh}^\mathrm{int,kin}(k,z)
\left[\Sigma_{hh}^\mathrm{kin,kin}(k,z)\right]^{-1}\ri\Omega_{hj}^\mathrm{kin}(k)$.
We can separate real and imaginary parts in memory functions~(\ref{math/3.60}) and~(\ref{math/3.61}) as follows: $\Sigma_{jj}(k,z)=\Sigma^{\prime}_{jj}(k,\omega)+\ri\Sigma^{\prime\prime}_{jj}(k,\omega)$ and
$\Pi_{hh}(k,z)=\Pi^{\prime}_{hh}(k,\omega)+\ri\Pi^{\prime\prime}_{hh}(k,\omega)$.
Herewith, the contributions from the hard sphere dynamics with typical
spatial-temporal scale $\tau_0\rightarrow+0$, $\Delta
r_0\rightarrow+0$ are separated in the transport kernel
$\varphi_{nn}(\vec{k};\vec{p},\vec{p}';t,t')$ only in the first
term in the right-hand side of elements~(\ref{math/3.47}) and hence in~(\ref{math/3.60}). After these transformations we can obtain a spectrum of collective
excitations in the hydrodynamic limit $k\sigma\ll1$: \textbf{
heat mode} $z_\mathrm{H}(k)=-D_\mathrm{T}k^2$, where
        $D_\mathrm{T}$ is the thermal diffusivity coefficient for the system with the
        potential of interaction~(\ref{math/3.48}). It has the following
        structure: $D_\mathrm{T}=D_\mathrm{TE}+D_\mathrm{T}^\mathrm{l}$,
        $D_\mathrm{T}^\mathrm{l}$ is determined through the corresponding
        elements~(\ref{math/3.47}) of matrix of transport
        kernels~(\ref{math/3.46}). $D_\mathrm{T}^\mathrm{l}=\frac{\lambda^\mathrm{l}}{nmc_{p}}$, $c_{p}$ is a heat capacity at constant
        pressure, $\lambda^\mathrm{l}$ is the heat conductivity coefficient
        in the hydrodynamic limit: $\lambda^\mathrm{l}=\lim_{k\rightarrow
        0,\omega\rightarrow 0}\lambda^\mathrm{l}(k,\omega)$. $\lambda^\mathrm{l}(k,\omega)$ is the generalized
        heat conductivity coefficient defined via elements of the
        matrix~(\ref{math/3.46}):
        $\lambda^\mathrm{l}(k,\omega)=\frac{c_{V}(k)}{k_\mathrm{B}\beta^{2}}\frac{1}{k^{2}}
        \Pi_{hh}^{\prime\prime}(k,\omega)$,
        where $c_{V}(k)$ is the generalized heat capacity at constant volume
        dependent on the wave vector $\vec{k}$;
\textbf{ two sound modes} $z_{\pm}(k)=\pm\,
        \ri ck-\Gamma k^2$, where $\Gamma$ is the sound damping and
        $c=\frac{c_{p}}{c_{V}\beta m S(0)}$ is the sound velocity in the system with the potential of interaction~(\ref{math/3.48}), $S(0)=S(k=0)$, $S(k)$ is a static structure factor of the system with potential~(\ref{math/3.48}).
        $\Gamma=\frac{1}{2}({c_{p}}/{c_{V}}-1)D_\mathrm{T}+\frac{1}{2}\eta^\mathrm{L}$, where $c_{V}=c_{V}(k=0)$,
        $\eta^\mathrm{L}=\left(\frac{4}{3}\eta^{\perp}+\eta^\mathrm{b}\right)/mn$ is the longitudinal viscosity defined via the bulk viscosity $\eta^\mathrm{b}=\eta^\mathrm{b}_\mathrm{E}+\eta^\mathrm{b}_\mathrm{l}$ and the shear viscosity
        $\eta^{\perp}=\eta^{\perp}_\mathrm{E}+\eta^{\perp}_\mathrm{l}$ coefficients. $\eta^{\perp}_\mathrm{E}$ is the shear viscosity in Enskog theory,
        and $\eta^{\perp}_\mathrm{l}$ is calculated in the hydrodynamic limit
        $\eta^{\perp}_\mathrm{l}=\lim_{k\rightarrow 0,\omega\rightarrow 0}\eta^{\perp}_\mathrm{l}(k,\omega)$. $\eta^{\perp}_\mathrm{l}(k,\omega)$
        is the generalized shear viscosity coefficient defined via elements of the matrix~(\ref{math/3.46})
         $\eta^{\perp}_\mathrm{l}(k,\omega)=\frac{mn}{\beta}\frac{1}{k^{2}}
         {\Sigma^{\prime\prime\perp}_{jj}}(k,\omega)$.
         $\Sigma^{\perp}_{jj}(k,z)$ is the transverse component of the generalized transport kernel $\Sigma_{jj}(k,z)$, where the wave vector $\vec{k}$ is directed along the $0Z$ axis.
        The longitudinal  viscosity coefficient $\eta^\mathrm{L}_\mathrm{l}$
        is calculated in the hydrodynamic limit $\eta^\mathrm{L}_\mathrm{l}=\lim_{k\rightarrow 0,\omega\rightarrow 0}\eta^\mathrm{L}_\mathrm{l}(k,\omega)$,
        where $\eta^\mathrm{L}_\mathrm{l}(k,\omega)$ is the generalized longitudinal viscosity coefficient  defined
        via longitudinal components of the generalized transport kernel
        $\Sigma_{jj}(k,z)$: $\eta^\mathrm{L}_\mathrm{l}(k,\omega)=\frac{mn}{\beta}\frac{1}{k^{2}}
        {\Sigma^{\prime\prime\mathrm{L}}_{jj}}(k,\omega)$;
\textbf{ two shear modes} with the eigenvalues given by
        $z_{\nu}(k)=-\nu k^2$.
        $\nu=\nu_\mathrm{E}+\nu_\mathrm{l}$ is the kinematic viscosity $\nu=\eta^{\perp}/nm$ for the system with the potential of interaction~(\ref{math/3.48}).
        Here, $\nu_l$ is a contribution determined by the
        corresponding elements~(\ref{math/3.47}) of the matrix
        of transport kernels~(\ref{math/3.46}).
In the limit $k\sigma\gg1$, we obtain a
\textbf{ diffusion mode}, with the eigenvalue
         $z_\mathrm{D}(k)=-D_\mathrm{E}k^2$, which is the same as in the Enskog theory.

As we can see from the above expressions, presence of the
long-range part in the potential of interaction entails a
renormalization of all the damping coefficients in the collective modes
spectrum. In particular, contributions related to long-range
potential appear in heat and sound modes as well as in shear modes.
Nevertheless, diffusion mode remains unchanged.

\section{Conclusions}

In this brief report within the framework of consistent
description of kinetic and hydrodynamic processes we considered a
set of kinetic equations for the potential of interaction of the
system presented by the sum of hard spheres potential
$\Phi^\mathrm{hs}(|\vec{r}_{ij}|)$ and a certain smooth one
$\Phi^\mathrm{l}(|\vec{r}_{ij}|)$. In this case, we separated the
Enskog-Boltzmann  collision integral describing a collision
dynamics at short distances from the collision integral of the
kinetic equation for the nonequilibrium distribution function.
Applying the procedure of projecting onto the moments of the
nonequilibrium distribution function to the
equations~(\ref{math/2.12}),~(\ref{math/3.49}) we obtain a set of
equations for hydrodynamic variables. Based on this set of equations
a spectrum of collective excitations was obtained in the limits
$k\sigma\ll1$ and $k\sigma\gg1$. We showed that, besides the
contribution from the hard spheres potential, all hydrodynamic modes contain contributions from the long-range part of potential. These contributions make the
damping coefficients closer to the ones known from the hydrodynamic
theory. Here, we formally presented the contribution from the
long-ranged part of potential, since the latter, for example the
Coulomb one, will contribute into the transport kernels~(\ref{math/2.14}).
Moreover, we can separate the linearized
Landau-like collision integral describing pair collisions in
$\varphi_{nn}(\vec{k};\vec{p},\vec{p}';t,t')$, while
$\varphi_{hn}(\vec{k};\vec{p};t,t')$,
$\varphi_{nh}(\vec{k};\vec{p};t,t')$,
$\varphi_{hh}(\vec{k};\vec{p};t,t')$ take into account collective
Coulombic interactions. Evidently, calculation of the elements
(\ref{math/3.47}) of matrix $\tilde{\Pi}(\vec{k};z)$ will depend
on the model of time dependence (exponential, Gaussian etc.) for
transport kernels (\ref{math/2.14}). When a spectrum of
collective excitations is known, a whole set of time correlation
functions can be investigated. In particular, it makes possible to
investigate the behaviour of the dynamic structure factor and, in the
case of potential~(\ref{math/3.48}), to separate a contributions
from the hard spheres potential and the long-range part of
potential in it.


\ukrainianpart

\title{До проблеми узгодженого опису кінетичних та гідродинамічних
процесів у густих газах та рідинах: \\ спектр колективних збуджень}
\author{Б.Б. Марків, І.П. Омелян, М.В. Токарчук}

\address{Інститут фізики конденсованих систем НАН України,
   вул.~І.~Свєнціцького, 1, 79011 Львів, Україна}

\makeukrtitle

\begin{abstract}
\tolerance=3000%
На основі отриманих раніше узагальнених немарківських рівнянь для
нерівноважної одночастинкової функції розподілу та середнього
значення густини потенціальної частини ентальпії [Markiv~B.B.,
Omelyan~I.P, Tokarchuk~M.V., Condens. Matter Phys., 2010, \textbf{13}, 23005] досліджується спектр колективних збуджень, коли
потенціал взаємодії між частинками представлено сумою потенціалу
твердих сфер та деякого далекосяжного потенціалу.
\keywords кінетика, гідродинаміка, кінетичні рівняння,  функції пам'яті, колективні моди

\end{abstract}

\end{document}